# An efficient topology optimization based on multigrid assisted reanalysis for heat transfer problem

Jichao Yin[*]; Hu Wang[†]; Daozhen Guo; Shuhao Li;

*State Key Laboratory of Advanced Design and Manufacturing for Vehicle Body*, *Hunan University*, *Changsha*, *410082*, *P.R. China*

**Abstract**

To improve the computational efficiency of heat transfer topology optimization, a Multigrid Assisted Reanalysis (MGAR) method is proposed in this study. The MGAR not only significantly improves the computational efficiency, but also relieves the hardware burden, and thus can efficiently solve large-scale heat transfer topology optimization problems. In addition, a projection-based post-processing strategy is also proposed and integrated with a continuous density filtering strategy to successfully obtain smooth boundary while eliminating some small-sized features. Several 2D and 3D numerical examples demonstrate that the computational efficiency of the MGAR is close to or even higher than that of the MGCG with almost identical optimization results, moreover, the efficiency improvement in the 3D scenario is superior than that of the 2D scenario, which reveals the excellent potential of the MGAR to save computational cost for large-scale problems.

---

[*] First author. E-mail address: jcyin@hnu.edu.cn (J. Yin)
[†] Corresponding author. E-mail address: wanghu@hnu.edu.cn (H. Wang)

# 1. Introduction

Topology optimization as a state-of-the-art structural design method was proposed by Bendsøe and Kikuchi in 1988[1], and after decades of development, it has been widely used in aircraft, automobiles, and architectures. It is well known that computational cost of solving governing equations is a dominant computational burden as Degrees of Freedom (DOFs) increases, especially for 3D problems with more than millions of DOFs[2]. In addition, finite hardware resources are also one of the important reasons for topology optimization to overcome the limitation of problem scale. Therefore, prohibitive computational cost and huge hardware requirements are critical challenges in application of topology optimization[3]. These challenges have motivated many researchers in the past few years to propose high-performance computational methods and alleviate the dependence of topology optimization on hardware resources.

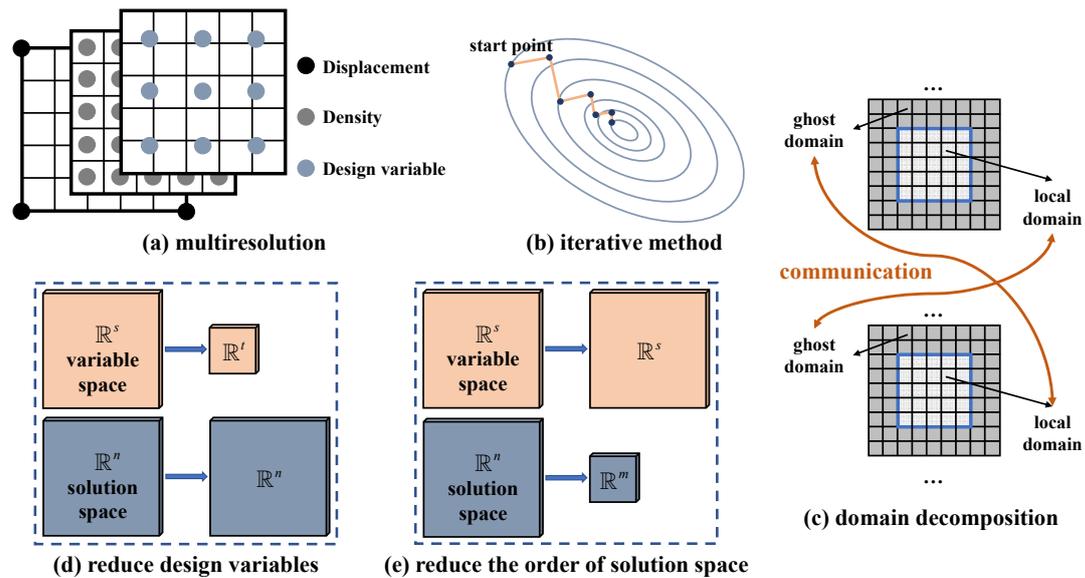

Figure 1 Different methods for reduction of the computational burden of topology optimization.

Methods to reduce the computational burden for topology optimization has been approached from various standpoints over the past few years as shown in Figure 1. One of the major researches is the Multiresolution Topology Optimization (MTOP)[4, 5], which reduced the computational burden by decoupling the optimization model into Finite Element (FE) mesh, density element mesh and the design variable mesh. In general, the resolution of the FE mesh is much lower than that of the density element mesh, and the resolution of the design variable mesh can be equal to or less than that of the density element mesh. However, the stiffness of models may be overestimated due to the Finite Element Analysis (FEA) performed on the coarse mesh, so Groen[6] *et al*. suggested employing higher-order element to improve the analysis accuracy. In addition, Liu[7] *et al*. integrated the MTOP into Moving Morphable Component (MMC) to improve computational efficiency while avoiding boundary effects, and thus achieve smooth representation of the boundary. Keshavarzzadeh[8] *et al*. improved the computational efficiency of reliability-based topology optimization and 3D topology optimization based on the MTOP while considering manufacturing uncertainty.

Other approaches for reduction of computational cost include iterative method such as preconditioned Krylov subspace methods with recycling[9], and parallel computation. The parallel technology integrated into the topology optimization framework can be divided into MPI[10, 11], CPU[12], GPU[13] parallel computation according to different implementation strategies. The strategy of domain decomposition[14] is a good choice for implementing parallel computing, and it should be noted that parallel computing strategies are often used in conjunction with iterative methods. Alternatively, topology optimization saves computational cost in the variable update phase by adaptively reducing the number of design variables based on their convergence history[15]. Reduced computational cost can also be achieved by adaptive mesh generation at the two-phase boundary, and a mixture of coarse and fine meshes significantly reduces the number of design variables[16]. Yoo[17] *et al*. proposed an adaptive Isosurface Variable Grouping (aIVG) strategy that is integrated into MTOP to reduce the computational burden of topology optimization.

However, as mentioned above solving is a dominant computational burden in the face of large-scale problems, the current research mostly focuses on dimensionality reduction of the solution space. Gogu[18] employed the displacement vectors to construct Reduced Order Model (ROM), and used Gram–Schmidt orthogonalization to ensure that the basis vectors are linearly independent. The original problem is recast into subspaces with smaller dimensions than the full model, thus achieving a reduction in computational effort. Yano[19] *et al*. proposed a projection-based ROM based on the trust-region method and posteriori error estimation theory, where the generalized trust-region constraints ensures that the optimization is globally convergent. Li[20] *et al*. successfully realized the application of ROM in dynamic topology optimization. Multigrid preconditioned Conjugate Gradients (MGCG)[21] solver can improve computational efficiency while ensuring the solution accuracy. The MGCG employed multigrid[22] (MG) as the preconditioner of the preconditioned conjugate gradient method, which can quickly eliminate the high-frequency and low-frequency errors, thus improve the convergence performance. The dimensionality of the linear system actually solved is related to the resolution of coarsest mesh. Xiao[23] *et al*. employed Principal Components Analysis (PCA) instead of Gram–Schmidt orthogonalization to construct ROM, and then coupled with MGCG to propose the on-the-fly multi-fidelity reduced model[24], which significantly saves computational cost.

The Combinatorial Approximation (CA) is also successfully implemented as a ROM in topology optimization, Amir[2] *et al*. integrated CA method into the topology optimization framework to efficiently solve small and medium scale problems of compliance minimization and compliant mechanism. Numerical results show that relatively rough approximations are acceptable. They also introduced the CA to solve the robust topology optimization of compliant mechanism design[25], and verified that the CA can improve the computational efficiency without influencing the outcome of the optimization process. Senne[26] *et al*. proposed combined CA and sequential piecewise linear programming method, which reduces the iterations required for topology optimization and further improves computational efficiency. Mo[27] *et al*. proposed an iterative reanalysis approximation for MMC to reduce iterations of topology optimization.

Heat transfer devices have been widely used in electronic devices, automotive industry, space shuttles and other fields[28]. To achieve better performance and cheaper cost, it is necessary to employ topology optimization to make full use of materials and improve thermal efficiency. Bendsøe[29] *et al*. implemented 2D steady-state heat transfer topology optimization based on the FE framework. Gersborg-Hansen[30] *et al*. successfully implemented heat transfer topology optimization using Finite Volume Method (FVM). Zeng[31] *et al*. employed a transient pseudo-3D fluid-thermal coupling model to perform topology optimization of heat sinks, which converted 3D analysis to 2D space to reduce design variables, significantly save computational cost. So far, there have been many related researches on topology optimization of heat transfer[32-34]. Jing[35, 36] *et al*. investigated the 2D heat transfer problem using the level set method, in which the Boundary Element Method (BEM) was used to perform physical analysis. Xia[37] *et al*. combined level set method and Bi-directional Evolutionary Structural Optimization (BESO) method to solve 2D heat transfer topology optimization. Jahangiry[38] *et al*. introduced Isogeometric Analysis (IGA) to solve 2D heat transfer topology optimization.

With the very recent exceptions of Yoo[17], to the knowledge of the authors, few of work has focused on the application of fast computational methods to solve topology optimization of heat transfer. Therefore, the purpose of this study is to propose an efficient topology optimization method for heat transfer topology optimization based on the framework of the CA method. To improve the efficiency of construction of Combined Approximation Reduced Model (CARM) and solve large scale problems, the MG method is employed. In addition, a projection-based post-processing strategy is proposed to obtain smooth boundary representations. This paper is organized as follows. The classical topology optimization method is briefly reviewed, the details of Multigrid Assisted Reanalysis (MGAR) method and post-processing strategy are proposed in Section 2. In Section 3, the MGAR is integrated into the topology optimization framework, and the influences of control parameters on the optimized process and outcomes are discussed. In Section 4, the proposed method is verified by several 2D and 3D examples in terms of efficiency improvement and quality of optimization results. Finally, some conclusions are given in Section 5.

## 2. Overview of theoretical formulation

A general heat transfer equation is given as

$$\rho c \frac{\partial T}{\partial t} - \nabla(k\nabla T) - Q = 0 \tag{1}$$

The study here is concerned with topology optimization of steady-state, isotropic, heat transfer exclusively, and the governing equation considering the Cartesian coordinate system is well known as[39]

$$\frac{\partial}{\partial x}(k_x \frac{\partial T}{\partial x}) + \frac{\partial}{\partial y}(k_y \frac{\partial T}{\partial y}) + \frac{\partial}{\partial z}(k_z \frac{\partial T}{\partial z}) + Q = 0 \tag{2}$$

where $k_{i,\ i=x,y,z}$ denote heat conductivities, $Q$ the heat source. Based on the isotropic assumption $k_x = k_y = k_z$, and Equation (2) can be approximated by means of a FE formulation as

$$\mathbf{K}t = q \tag{3}$$

where $\mathbf{K}$ denotes the conductivity matrix, $t$ the nodal temperature vector and $q$ the applied heat load vector.

### 2.1 Topology optimization formulation

With the modified Solid Isotropic Material interpolation with Penalization (SIMP) method, the conductivity $k$ is interpolated as

$$k = k_{\min} + (k_0 - k_{\min})\tilde{\rho}^p \tag{4}$$

where $k_0$ and $k_{\min}$ are heat conductivities associated with solids and voids ($k_0 \gg k_{\min}$), respectively. $p$ is the penalty factor which is usually 3. $\tilde{\rho}$ is physical density of the corresponding FE model. Employing SIMP method to transform the discrete design variables to continuous, the topology optimization formula is written as follows

$$\begin{aligned}
&\min: f(\rho) = t^T \mathbf{K} t \\
&\text{s.t.:} \mathbf{K}(\tilde{\rho})t = q \\
&\quad g_v(\rho) \leq 0 \\
&\quad 0 \leq \rho_e \leq 1, \quad e = 1, 2, ,, N
\end{aligned} \tag{5}$$

where $g_v(\boldsymbol{\rho})$ denotes volume constraint. It should be noted that physical density $\tilde{\boldsymbol{\rho}}$ is a function of design variables $\boldsymbol{\rho}$, such as when using density filter[40]

$$\tilde{\rho}_e = \frac{\sum_{i \in Ne} \omega(\boldsymbol{\rho}_i) v_i \rho_i}{\sum_{i \in Ne} \omega(\boldsymbol{\rho}_i) v_i} \tag{6}$$

with

$$Ne = \{ m \mid \|\boldsymbol{\rho}_m - \boldsymbol{\rho}_e\| \leq r \} \tag{7}$$

where $v_i$ is the volume of $i$-th element, $Ne$ denotes element set contained in the region centered on $e$-th element with radius $r$. The weighting function $\omega(\boldsymbol{\rho}_i)$ is a linearly decaying (cone-shape) function with the distance between elements as a variable.

$$\omega(\boldsymbol{\rho}_i) = \max(0, r - \|\boldsymbol{\rho}_i - \boldsymbol{\rho}_e\|) \tag{8}$$

The sensitivity of the objective function $f(\boldsymbol{\rho}) = \boldsymbol{t}^T \mathbf{K} \boldsymbol{t}$ with respect to design variable can be derived by the chain rule

$$\frac{\partial f(\boldsymbol{\rho})}{\partial \rho_e} = \sum_{i \in Ne} \frac{\partial f(\boldsymbol{\rho})}{\partial \tilde{\rho}_i} \frac{\partial \tilde{\rho}_i}{\partial \rho_e} \tag{9}$$

with

$$\frac{\partial f(\boldsymbol{\rho})}{\partial \tilde{\rho}_i} = -\boldsymbol{t}^T \frac{\partial \mathbf{K}}{\partial \tilde{\rho}_i} \boldsymbol{t} = -p(k_0 - k_{\min}) \tilde{\rho}_i^{p-1} \boldsymbol{t}^T \mathbf{k}_i^0 \boldsymbol{t} \tag{10}$$

and

$$\frac{\partial \tilde{\rho}_i}{\partial \rho_e} = \frac{\omega(\boldsymbol{\rho}_e) v_e}{\sum_{j \in Ni} \omega(\boldsymbol{\rho}_j) v_j} \tag{11}$$

## 2.2 Multigrid assisted reanalysis method

The variable $\Delta \mathbf{K} = \mathbf{K} - \mathbf{K}_0$ is introduced to describe the variation between the conductivity matrices corresponding to the current design cycle and the factorized, so the governing equation (3) can be rewritten as follows

$$(\mathbf{K}_0 + \Delta \mathbf{K}) \boldsymbol{t} = \boldsymbol{q} \tag{12}$$

where $\mathbf{K}_0$ is the factorized conductivity matrix. The recurrence formula of solution

$\mathbf{K}_0 \boldsymbol{t}^{(k)} = \boldsymbol{q} - \Delta\mathbf{K}\boldsymbol{t}^{(k-1)}$ is established by Equation (12), and the approximate nodal temperature vector can be expressed by employing the reference nodal temperature vector as

$$\boldsymbol{t} = \boldsymbol{t}_0 + (-\mathbf{K}_0^{-1}\Delta\mathbf{K})\boldsymbol{t}_0 + (-\mathbf{K}_0^{-1}\Delta\mathbf{K})^2 \boldsymbol{t}_0 + (-\mathbf{K}_0^{-1}\Delta\mathbf{K})^3 \boldsymbol{t}_0 + \ldots \tag{13}$$

where the reference nodal temperature vector $\boldsymbol{t}_0$ is obtained by solving the governing equation involving the factorized conductivity matrix $\mathbf{K}_0 \boldsymbol{t}_0 = \boldsymbol{q}$. It should be noted that the factorized conductivity matrix used to construct the approximate solution is updated dynamically during the optimization process rather than being fixed. When the accuracy of the approximate solution violates a given criterion, a new factorization of the conductivity matrix should be performed.

Equation (13) indicates that the nodal temperature vector is approximated by the linear combination of infinite terms, but its numerical implementation is impractical. Therefore, the CA employs the first $m$ terms of binomial series sequence[41] as basis vectors of CARM (**R**), and thus the approximate solution is obtained from a linear combination of the basis vectors.

$$\begin{aligned}\tilde{\boldsymbol{t}} &= y_1 \boldsymbol{t}_0 + y_2 (-\mathbf{K}_0^{-1}\Delta\mathbf{K})^1 \boldsymbol{t}_0 + \ldots + y_m (-\mathbf{K}_0^{-1}\Delta\mathbf{K})^{m-1} \boldsymbol{t}_0 \\ &= y_1 \boldsymbol{r}_1 + y_2 \boldsymbol{r}_2 + \ldots + y_m \boldsymbol{r}_m \\ &= \mathbf{R}\boldsymbol{y}\end{aligned} \tag{14}$$

with

$$\begin{cases} \boldsymbol{r}_1 = \boldsymbol{t}_0 \\ \mathbf{K}_0 \boldsymbol{r}_i = -\Delta\mathbf{K}\boldsymbol{r}_{i-1} \end{cases} \tag{15}$$

Substitute Equation (14) into Equation (3) to obtain the modified governing equation

$$\mathbf{K}\mathbf{R}\boldsymbol{y} = \boldsymbol{q} \tag{16}$$

However, the variation of design variables in topology optimization is global, especially for heat transfer problem with finger-like[29] results, the original structure varies commonly greatly. The response of the varied structure might not be well predicated by CA due to the large variation of conductivity matrix, thus dynamic factorization of the conductivity matrix and reconstruction of CARM are necessary. Therefore, computational cost of high-frequency factorization is prohibitive, moreover, the storage of the factorized dense matrix might be limited by hardware devices. To handle this problem, the MG method is introduced.

MG is a widely used multilevel iterative method for solving linear systems with a computational cost of $O(n)$[21, 42]. The keys to rapid convergence of MG can be summarized as: (*i*) the original FE mesh is coarsened sequentially into meshes with different resolution, and the adjacent meshes are connected using restriction and interpolation operators; (*ii*) high-frequency errors are eliminated using the smoother (damped Jacobi method $x = x + \omega \mathbf{D}^{-1}(b - \mathbf{A}x)$, where $\mathbf{D}$ is the diagonal of the coefficient matrix $\mathbf{A}$); (*iii*) lower-frequency errors are eliminated by solving the residual equation at the coarsest mesh. V-cycle as the most common MG strategy is applied in this work[21], and its pseudocode (coarsening level is 2) is given in Table 1.

Table 1 Algorithm implementation for V-cycle strategy.

| V-cycle with 2 layers: $x \leftarrow MG(\mathbf{A}, b, x, \mathbf{P})$ | | |
|---|---|---|
| 1: | Pre-smooth | $x \leftarrow smoother(\mathbf{A}, b, x)$ |
| 2: | Calculate residual | $res = \mathbf{P}^{\mathrm{T}}(b - \mathbf{A}x)$ |
| 3: | Coarse mesh correction | $\mathbf{A}^c x^c = res$ |
| 4: | Interpolation | $x = x + \mathbf{P}x^c$ |
| 5: | Post-smooth | $x \leftarrow smoother(\mathbf{A}, b, x)$ |

Employ the MG method to calculate the basis vectors of CARM, Equation (15) can be rewritten as

$$\begin{cases} r_1 = t_0 \\ r_i \leftarrow MG(\mathbf{K}_0, -\Delta \mathbf{K} r_{i-1}, r_{i-1}, \mathbf{P}) \end{cases} \quad (17)$$

Furthermore, to avoid singularity in the computational process, the PCA method is a good choice to make the basis vectors orthogonal.

$$\mathbf{R} = \tilde{\mathbf{R}} \Sigma \mathbf{V}^{\mathrm{T}} \quad (18)$$

Employing the desired $\tilde{\mathbf{R}}$ and multiplying both sides of Equation (16) by $\tilde{\mathbf{R}}^{\mathrm{T}}$

$$(\tilde{\mathbf{R}}^{\mathrm{T}} \mathbf{K} \tilde{\mathbf{R}}) y = \tilde{\mathbf{R}}^{\mathrm{T}} q \quad (19)$$

with

$$\tilde{t} = \tilde{R}y \tag{20}$$

Equation (19) can be simplified to

$$K_{\tilde{R}} y = q_{\tilde{R}} \tag{21}$$

where $K_{\tilde{R}} \in \mathbb{R}^{m \times m}$, $m$ is specified by the users, usually less than ten, far less than the dimension of the governing equation, which is usually on the order of millions.

## 2.3 Post-processing strategy for boundary smoothing

Most of results of topology optimization about steady-state heat transfer problem are finger-like, and their numerous small-sized features are not conducive to industrial manufacturing (as shown in Figure 2(a)). The projection-based size control strategy is suggested to remove small-sized features. Common ones include original Heaviside function, modified Heaviside function, volume preserving Heaviside function and hyperbolic tangent function, the detailed introduction of projection functions are given in references [43, 44]. The design domain of the optimized model is given in Figure 6, and the boundary conditions involved are consistent with those in Section 3.2.

The filter radius is set in accordance with the literature[43], and the optimization results based on different projection functions are shown in Figure 2, it should be noted that all projection strategies are used in conjunction with density filter. The optimized results of employing projection methods tend to be consistent with pure density filter in terms of the distribution of the "trunk", but less small-sized features appear in projection methods. Among them, the optimization objective of employing original Heaviside is the worst, and too many gray elements exist means that original Heaviside is poorly suppressed for low-density elements. The gray elements in Figure 2(c), (d), and (e) are significantly reduced, however, the optimization process is prone to instability when the penalty factor of the projection method is too large. Moreover, the frequency of reconstructing the CARM increases due to the non-continuous increase of the penalty factor, which will generate more computational burden especially for factorizing matrix.

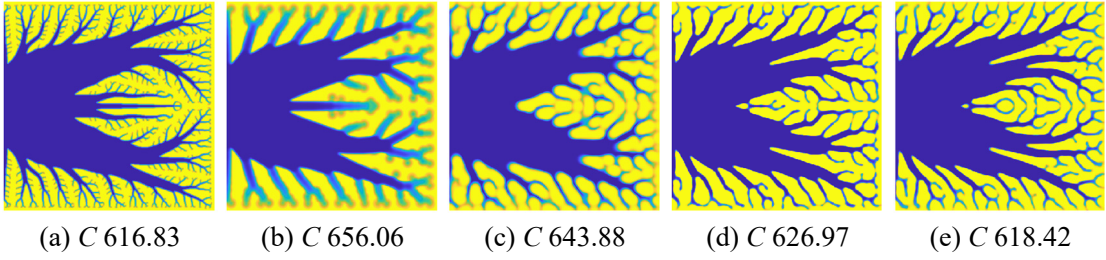

(a) *C* 616.83    (b) *C* 656.06    (c) *C* 643.88    (d) *C* 626.97    (e) *C* 618.42

Figure 2 Optimization results after 300 design cycles with different filters. (**a**) employed pure density filter; (**b**) employed original Heaviside; (**c**) employed modified Heaviside; (**d**) employed volume preserving Heaviside; and (**e**) employed hyperbolic tangent. The filter radii were set to 3 in (**a**) and 1.4L/50 in (**b-e**).

For the above reasons, a continuous density filtering strategy is suggested rather than projection-based Heaviside functions, which can significantly reduce small-sized features and ensure the optimization process is as smooth and stable as possible.

$$r^{(k)} = \max(r_{\min}, \eta r^{(k-1)}) \tag{22}$$

with

$$\eta = e^{\frac{\ln(r_{\min}) - \ln(\alpha L)}{lp}} \tag{23}$$

where $k$ is the index of design cycles, $r_{\min}$ is the minimum filter radius allowed, the constant factor $\alpha$ is used to calculate the initial filter radius $r_0 = \alpha L$, and $lp$ is the number of design cycles required to reduce the radius to $r_{\min}$. If not otherwise specified, $\alpha$ is set to 1.4/50 and $lp$ is set to 5/6 of the maximum design cycle in this work.

As shown in Figure 4(a) with $r_{\min}$ of 3, the application of the continuous density filtering strategy eliminates most of the small-sized features relative to simple density filter, and its optimization objective is at the same level as the Heaviside functions, or even better than most of them. Moreover, the gray elements are suppressed because the filter radius is kept at a minimum value $r_{\min}$ at the final stage of optimization. However, zigzag boundary is not feasible for industrial manufacturing, to improve the manufacturability of the optimization results, a post-processing strategy is proposed to achieve a smooth representation of the boundary.

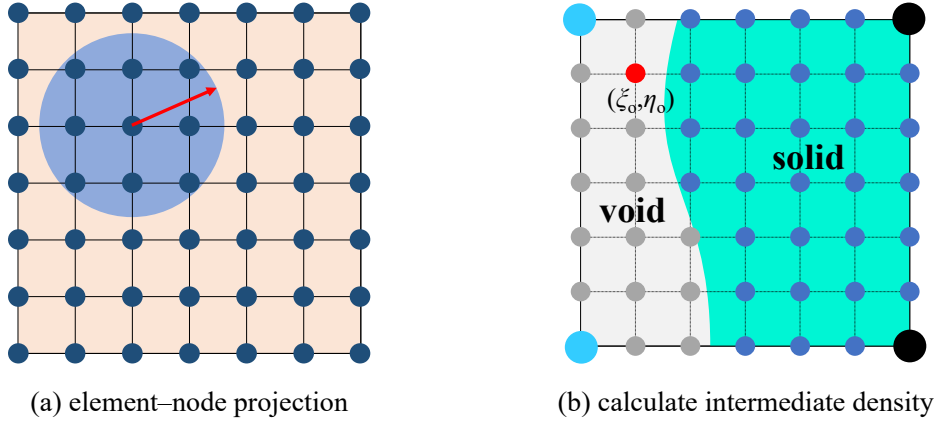

(a) element–node projection  (b) calculate intermediate density

Figure 3 Projection-based boundary smoothing strategy.

As shown in Figure 3(a), the element sensitivity corresponding to the density field are calculated, and then the element sensitivity is projected onto the node in a manner analogous to density filter. Here, inspired by the idea of the level set method[45], the distribution of material in the design domain is determined by the value of the nodal sensitivities (*ns*).

$$\rho_e = \begin{cases} 1, & ns_i > ns_{level} \\ 0, & ns_i < ns_{level} \\ ev, & otherwise \end{cases}, \quad i \in C_e \tag{24}$$

where $C_e$ denotes the all nodes associated with *e*-th element. The criterion for selecting $ns_{level}$ is to ensure the consistency of volume fraction before and after post-processing. It should be emphasized that the volume fraction varies monotonically with respect to the choice of $ns_{level}$, so $ns_{level}$ can be calculated employing a one-dimensional search method such as bisection. To determine the element densities *ev* that mixes the node sensitivities on both sides of the $ns_{level}$, such elements are again discrete and use interpolation function to calculate the level set function (LSF) value of new nodes (the smaller dots in Figure 3(b)).

$$ns_{new}(o) = \sum_{i \in C_e} N_i(o) ns_e^i \tag{25}$$

with

$$N_i(o) = \frac{(1+\xi_o \xi_i)(1+\eta_o \eta_i)}{4} \tag{26}$$

where $ns_e^i$ denotes LSF value at the *i*-th original node (large dots) of *e*-th element, $N_i(o)$ the interpolation function, $(\xi_i, \eta_i)$ and $(\xi_o, \eta_o)$ the local coordinates of the original nodes and the new nodes in Figure 3(b), respectively. The density of such elements is determined by the percentage of node sensitivity over $ns_{level}$, and the smooth boundaries can be represented by node sensitivities close to $ns_{level}$.

$$ev = \frac{|\{(\xi_m, \eta_m) | ns_m > ns_{level}, m \in C_{total,e}\}|}{|C_{total,e}|} \tag{27}$$

where the $|\cdot|$ denotes the number of elements in the statistical set, $C_{total,e}$ the node set corresponding to $e$-th element, which including original nodes and new nodes.

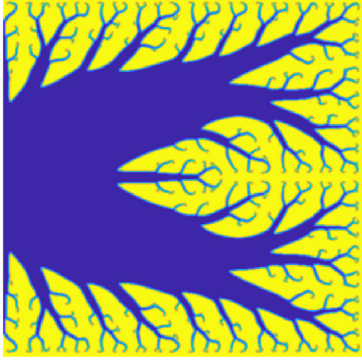 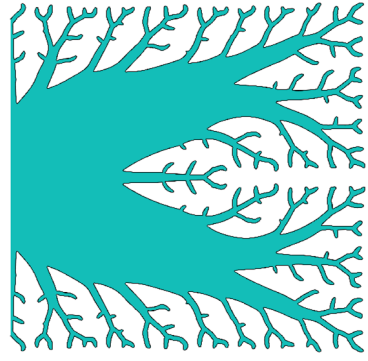

(a) continuous density filtering  (b) boundary smoothing

Figure 4 Post-processing results after 300 design cycles with continuous density filtering. The optimization objectives are 619.34(**a**) and 645.42(**b**).

The result obtained by employing the post-processing strategy is shown in Figure 4(b). The overall trend of material distribution remains consistent relative to the structure before smoothing, and zigzag boundaries are eliminated. In addition, a positive by-product is that small-sized features are further eliminated. The optimization objective after post-processing is slightly increased, but this is acceptable to improve the manufacturability of the optimization results.

## 3. MGAR-based topology optimization process

The details of the MGAR and post-processing strategy have been introduced in the previous section. In this section, the MGAR method and post-processing strategy are integrated in the topology optimization framework, and compared with MGCG in terms of computational efficiency.

## 3.1 Optimization process with post-processing

To represent the process of optimization procedure clearly, the flowchart is shown in Figure 5. In contrast to the classical SIMP framework[29], the solution module is replaced by the MGAR (red wireframe in Figure 5), and the post-processing module is placed after the optimization process satisfies the stopping criteria. In addition, the MGCG is used as an auxiliary solver when approximate solutions obtained by CARM does not satisfy the requirements.

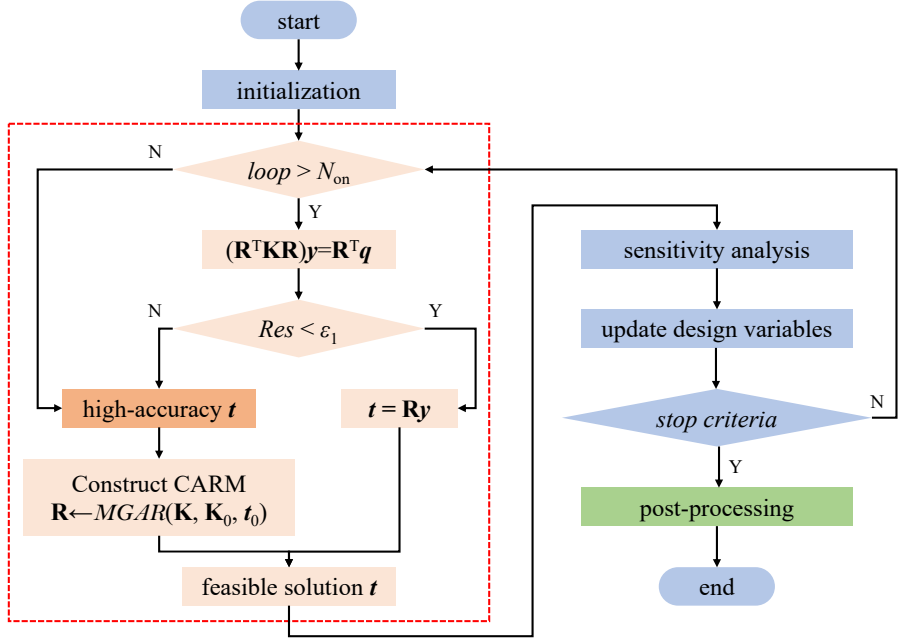

Figure 5 Flowchart of MGAR-based topology optimization.

It is well known that design variables vary greatly in the early stage of topology optimization, and the reanalysis method is difficult to be used for such drastic structural modifications. The approximate solution obtained by MGAR at this stage does not satisfy the accuracy requirement and corresponding operations should be useless. Therefore, activation parameters $N_{on}$ is proposed to control the activation of MGAR. When the design cycle is less than $N_{on}$, the auxiliary solver MGCG is employed; otherwise, the low-dimensional linear system based MGAR is solved. It should be noted that the initial solution $t_0$ for constructing the CARM is the approximate nodal temperature vector from the previous iteration.

The approximate solution obtained based on MGAR needs to determine whether it satisfies the accuracy requirements. If the residual norm *Res* is less than the threshold $\varepsilon_1$, the approximate solution obtained by MGAR is considered feasible. Otherwise, it is considered invalid, so the auxiliary solver MGCG is employed to calculate approximate solutions and the CARM should be reconstructed. For simplicity of distinction, the threshold $\varepsilon_1$ is referred to as the reconstruction criterion.

$$Res = \frac{\|\mathbf{KR}y - q\|}{\|q\|} \qquad (28)$$

Moreover, it should be emphasized that the post-processing is carried out only once after the completion of the iteration, rather than being used repeatedly during the optimization process, so the computational cost occupied by the post-processing strategy can be neglected and the influence of post-processing operations on the optimization process can be avoided.

## 3.2 Efficiency with different MG coarsening levels

Both MGAR and MGCG involve the application of MG, where the coarsening level of MG significantly influences the solution efficiency. Therefore, the computational efficiency with different coarsening levels is discussed here for the example with volume fraction of 0.5 and minimum filter radius $r_{min}$ of 3. The design domain is the orange area of the square shown in Figure 6 with the internal heat source per unit volume is set to 1e-4. The design domain is discretized with a 480 × 480 mesh for the FEA of temperature field. The conductivity $k_0 = 1$ for solid material and $k_{min} = 1e-3$ for void material to avoid singularity. The grey area indicates that the boundary is adiabatic, and the blue boundary (1/10 of side length) in the middle of the left side denotes the Dirichlet boundary condition with T = 0.

The CARM consists of two basis vectors, the reconstruction criterion $\varepsilon_1$ is set 0.5, and the activation parameter $N_{on}$ is set to 2/15 of the maximum iteration. The convergence condition of MGCG is that the solution accuracy less than a small value $\varepsilon_2$ or conjugate gradient (CG) iterations reach the maximum $cg_{max}$. Here refer to the settings of [21], $\varepsilon_2 = $ 1e-6, $cg_{max} = $ 200, which are consistent for both pure MGCG and MGCG involved in MGAR. First, the coarsening level are set to 2, 3 and 4 respectively; Second, the average computational cost per iteration step of the pure MGCG with coarsening level of 3 is used as a unit cost measure. The normalized computational cost of the iterative process, as well as the efficiency improvement of MGAR relative to pure MGCG is accounted for and graphically shown in Figure 7.

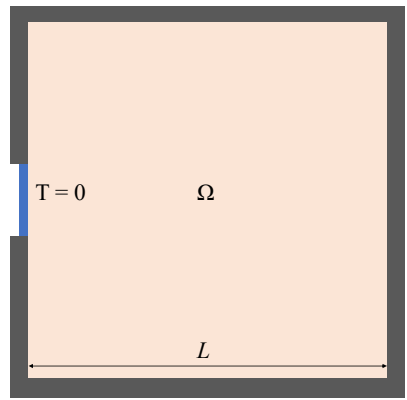

Figure 6 Design domain and boundary conditions for steady-state heat transfer problem.

The coarsening level has a significant effect on both MGAR and pure MGCG, and is more prominent in the latter. The total computational cost of MGAR is almost multiplied as the coarsening level increases, and rises even more for pure MGCG. In addition, the cumulative cost of pure MGCG maintains linear growth throughout the optimization process, but the growth rate of the cumulative cost of MGAR slows down when the design cycle is greater than $N_{on}$. The lower the coarsening level, the more significant this phenomenon is, which also indirectly indicates that MGAR is more efficient than MGCG under the current conditions. However, it is unreasonable to pursue higher coarsening levels, the gains from increasing the coarsening level will gradually weaken. On the other hand, the FE mesh discretization of physical model is limited by the coarsening level, the number of axial elements must be a multiple of $2^{nl-1}$ ($nl$ denotes coarsening level).

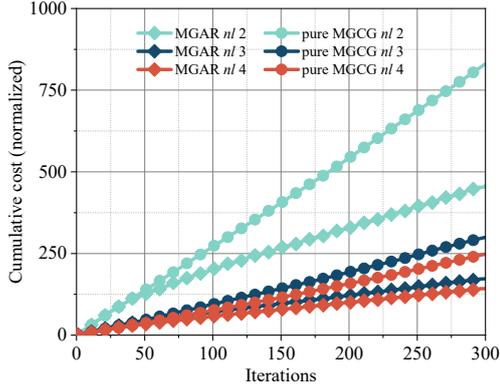 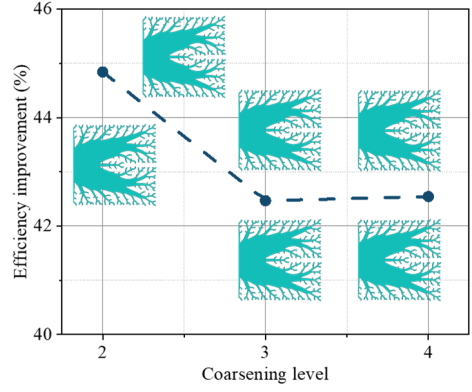

(a) normalized cumulative cost curve  (b) Efficiency improvement

Figure 7 Comparison of computational efficiency with different coarsening levels. Optimization results on the upper of the curve are obtained by MGAR in (**b**).

Figure 7(b) illustrates the efficiency improvement and optimized structures of MGAR relative to pure MGCG at different coarsening levels. Although the efficiency improvement performs best with a coarsening level of 2, the total computational cost is most expensive than in other cases, which is clearly not an appropriate choice. The efficiency improvements for coarsening levels of 3 and 4 are close, the mesh discretization has the strongest dependence on the number of axial elements when the coarsening level is 4. In contrast, the coarsening level of 3 significantly weakens the dependence on mesh discretization and ensures the advantage of efficiency improvement, thus a coarsening level of 3 is most appropriate. In addition to improving efficiency, the optimized structures of MGAR and MGCG are almost identical in details, and the relative differences of the optimization objective are controlled to a low level, with a maximum value of only 0.095%.

## 3.3 Efficiency with relaxed MGCG convergence criteria

The coarsening level of 3 is a good choice to trade-off the geometric dependence of mesh discretization and the cumulative computational cost, so this coarsening level is set by default in the following tests. It should be noted that two criteria were applied in this work, the reconstruction criterion $\varepsilon_1$ for controlling the dynamic reconstruction of CARM and $\varepsilon_2$ for controlling the accuracy of the MGCG solver. In general, the former is larger than the latter, which means that the accuracy requirements of MGCG are more stringent. For this reason, $\varepsilon_2$ is set to 0.5, 1e-3 and 1e-6, respectively, to capture the efficiency improvement of MGAR in more detail.

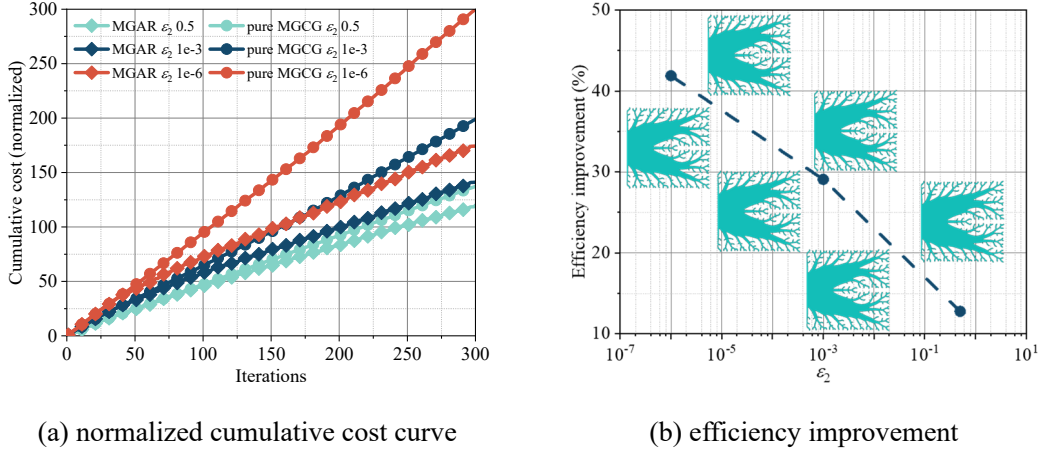

(a) normalized cumulative cost curve    (b) efficiency improvement

Figure 8 Normalized cumulative cost curve with different solution accuracy, the average cost per iteration of pure MGCG with a coarsening level of 3 is taken as a measure.

The rest of the parameter settings are consistent with Section 3.2 except for the setting of $\varepsilon_2$. Relaxing $\varepsilon_2$ means that fewer CG iterations are required for MGCG to achieve the required accuracy, thus reducing the computational cost. Therefore, the accuracy requirement significantly influences the computational cost, especially for pure MGCG. With the relaxation of $\varepsilon_2$, the reduction of the total computational cost of pure MGCG is greater than that of MGAR, and the gap between them is becoming smaller. The reason for this is the relatively low frequency of MGAR invokes auxiliary MGCG solver. Nevertheless, the MGAR still has advantages over pure MGCG when the accuracy requirement $\varepsilon_2$ is consistent with the criterion for reconstructing CARM ($\varepsilon_1 = 0.5$), the efficiency can be improved by 12.77%. Although the efficiency improvement is sensitive to $\varepsilon_2$, the optimization results are almost unaffected by it at the current settings. No matter between different accuracy requirements $\varepsilon_2$ or between MGAR and MGCG, the structural topologies remain consistent in details and the maximum difference between optimization objectives is controlled to within 0.1%.

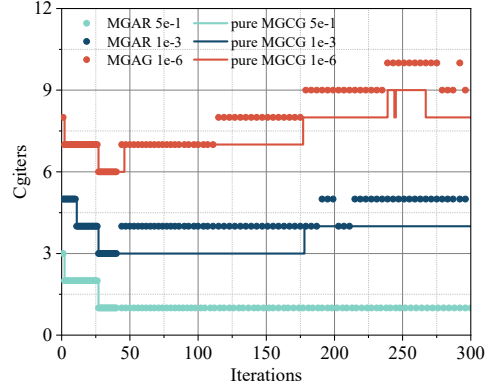

Figure 9 The number of CG iterations performed by MGCG per design cycle.

To further clarify the reason for the difference in efficiency improvement, the CG iterations per design cycle is counted in Figure 9. It should be noted that MGCG is called intermittently in MGAR, and the interval between adjacent calls is the number of iterations between the two maximum residuals in Figure 10. Under the same accuracy requirement, the CG iterations required for MGCG called in MGAR are always equal to or higher than pure MGCG. Therefore, MGCG playing the role of an auxiliary solver does not contribute to the efficiency improvement in the MGAR-based topology optimization, and even causes more computational cost. If the influence of the auxiliary MGCG solver is excluded, the actual efficiency improvement of MGAR should be higher. In addition, the residual of MGAR is typically oscillating between $\varepsilon_1$ and $\varepsilon_2$, and the oscillating frequency is relatively low in the middle and late iterations. The accuracy achieved by MGAR is worse than that achieved by pure MGCG with $\varepsilon_2 = 0.5$, which is attributed to the fact that MGCG only performs one CG iteration to satisfy the accuracy requirements, the pure MGCG provides higher quality initial solutions and thus achieves better approximate accuracy. The MGCG requires more CG iterations when the accuracy requirement is more stringent, the MGAR can achieve the same accuracy as pure MGCG in some design cycles. Moreover, the frequency of MGAR calling auxiliary MGCG solver is not sensitive to $\varepsilon_2$, the 114 MGCG evaluations for $\varepsilon_2 = 0.5$ and 108 MGCG evaluations for the remaining two settings.

However, it should be emphasized that MGAR and MGCG were not in competition, but in cooperation. The test results show that MGAR is as efficient as pure

MGCG, and in some cases even better. Although the residuals are influenced by relaxed MGCG convergence criteria, this is negligible for the final optimization results.

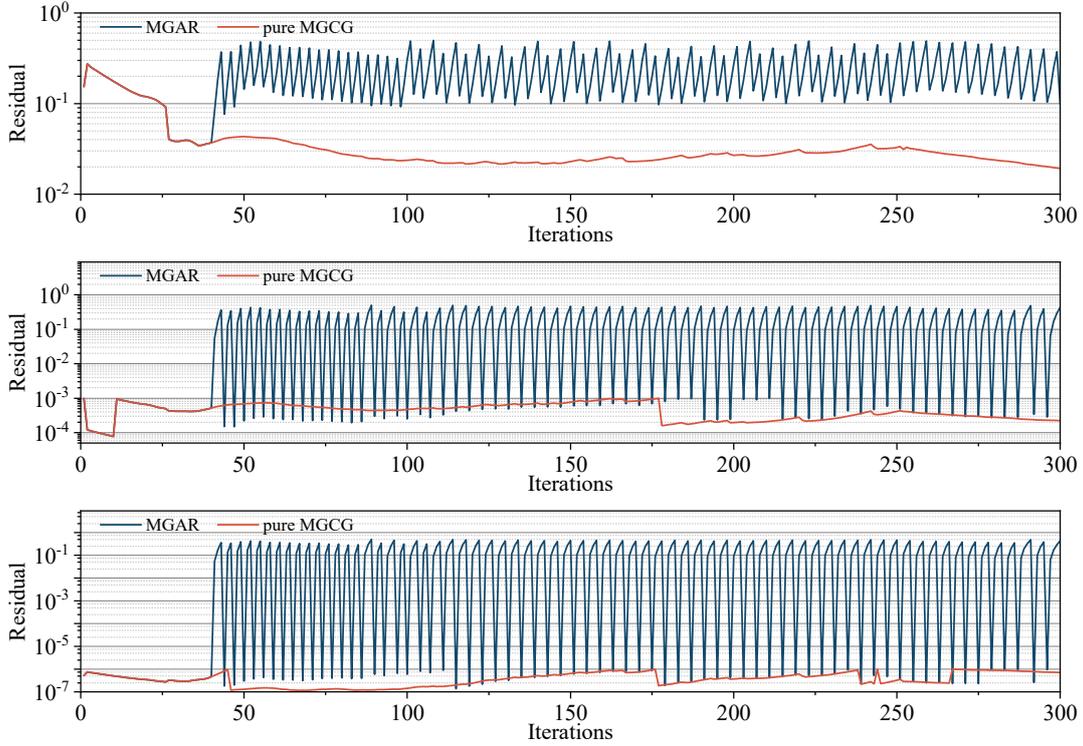

Figure 10 The residual curve of the iterative process, the accuracy of MGCG is 0.5, 1e-3, 1e-6 from top to bottom.

## • 4. Numerical examples

To study the competence of MGAR for different scenarios, the 2D uniform heat source models with different FE meshes and 2D non-uniform heat source models are discussed. To extend the scope of MGAR applications, the 3D uniform heat source model is likewise discussed. If not specified, all tests have a reconstruction criterion $\varepsilon_1$ of 0.5, a coarsening level $nl$ of 3, an accuracy requirement $\varepsilon_2$ of 1e-6 and a maximum number of CG iterations $cg_{max}$ of 200.

### 4.1 2D uniform heat source models

The geometry, boundary conditions and material properties are set the same as in Section 3.2, but the scale of the FE mesh is different. In ascending order of mesh scale as: Model 1 ($L = 720$), Model 2 ($L = 1560$), Model 3 ($L = 2040$), Model 4 ($L = 2480$), the internal heat source per unit volume is set to 1e-4. The CARM consists of two basis vectors, the activation parameter $N_{on}$ is 2/15 of the maximum iteration, the desired volume fraction is 0.5 and the minimum filter radius $r_{min}$ is 3.

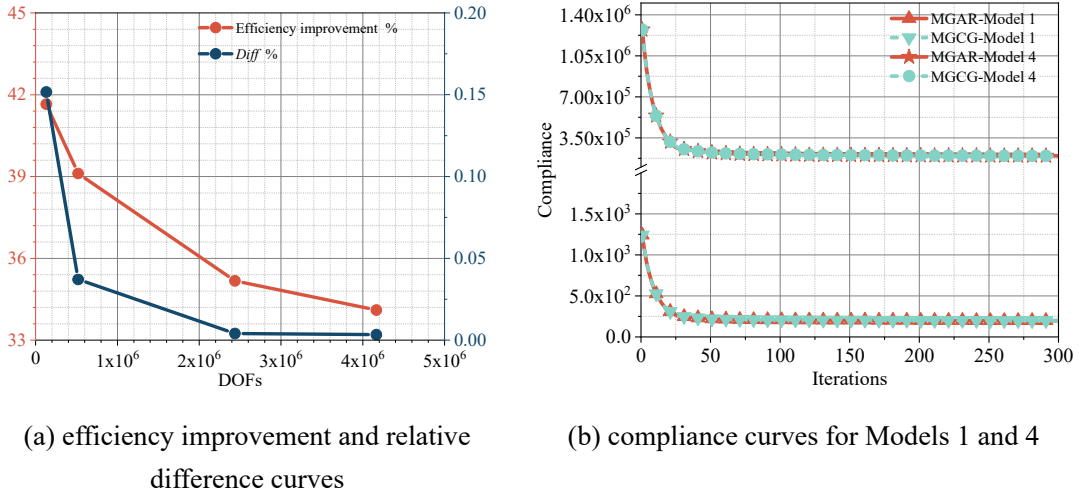

(a) efficiency improvement and relative difference curves

(b) compliance curves for Models 1 and 4

Figure 11 Efficiency improvement and relative difference in optimization objectives.

The comparison of MGAR with respect to pure MGCG in terms of efficiency improvement as well as relative difference of the optimization objectives is given in Figure 11(a) and Table 2. The authors found that the efficiency improvement of MGAR remain attractive when dealing with large-scale problems, although showing a decaying trend as the problem scale increases, but gradually saturates and can be expected to eventually stabilize at a respectable value. In the current test, the worst efficiency improvement can still reach 34.10% when the problem scale is about 4.2 million DOFs. The total computational cost of MGAR in Model 1 is used as a unit measure, the total computational cost of pure MGCG is increasing despite the fact that the efficiency improvement shows a decaying trend. However, the statistics in Table 2 show that the growth rate of the computational cost of pure MGCG far exceeds the decay rate of the efficiency improvement, the savings for large-scale problems are more significant than for small-scale problems, so the actual cost savings increase with the problem size despite the reduced efficiency improvement, the MGAR shows great potential when solving large-scale problems.

The performance of saving computational cost is guaranteed while the consistency of optimization results relative to pure MGCG is also achieved, the difference of optimization objectives as shown in Figure 11(a) is much smaller. The maximum difference up to 0.15% and the minimum difference is only 0.0033%, which can be negligible. Not only that, investigating the objective curves for models 1 and 4 given in Figure 11(b) reveals that the iterative processes of MGAR and pure MGCG are almost exactly coincident and vary smoothly. This further verifies that MGAR not only dominates in computational efficiency, but also maintains excellent optimization stability and accuracy during the whole optimization process.

Table 2 Comparison of computational results with different FE mesh scale.

|  | Model 1 (DOFs 130,321) | Model 2 (DOFs 519,841) | Model 3 (DOFs 2,436,721) | Model 4 (DOFs 4,165,681) |
|---|---|---|---|---|
| Cost of MGAR | 1 | 5.58 | 42.25 | 102.81 |
| Cost of pure MGCG | 1.71 | 9.16 | 65.17 | 156.01 |
| **Savings** | **0.71** | **3.58** | **22.92** | **53.20** |
| MGCG evaluations | 107 | 115 | 125 | 127 |
| Efficiency improvement | 41.65% | 39.11% | 35.18% | 34.10% |
| Relative difference | 0.15% | 0.037% | 0.0040% | 0.0033% |

According to Table 2, a total of 107 MGCG evaluations were required for Model 1 and 127 MGCG evaluations for Model 4, the frequency of calling MGCG increased by 18.7%. Solving large-scale problems is accompanied by an increase in MGCG evaluation, while efficiency improvement presents a downward trend. And considering that MGCG is more expensive than MGAR in a single iteration, it can be inferred that the MGCG evaluations is one of the important reasons influencing the efficiency improvement. In addition, MGCG evaluations is close to the performance of efficiency improvement in terms of magnitude of variation and can foresee that both will eventually stabilize. Even if the efficiency improvement is already saturated, the cost savings will be even more significant due to the fact that the cost of the original problem is increasing, MGAR has a great potential for solving large-scale and even super-scale heat transfer topology optimization.

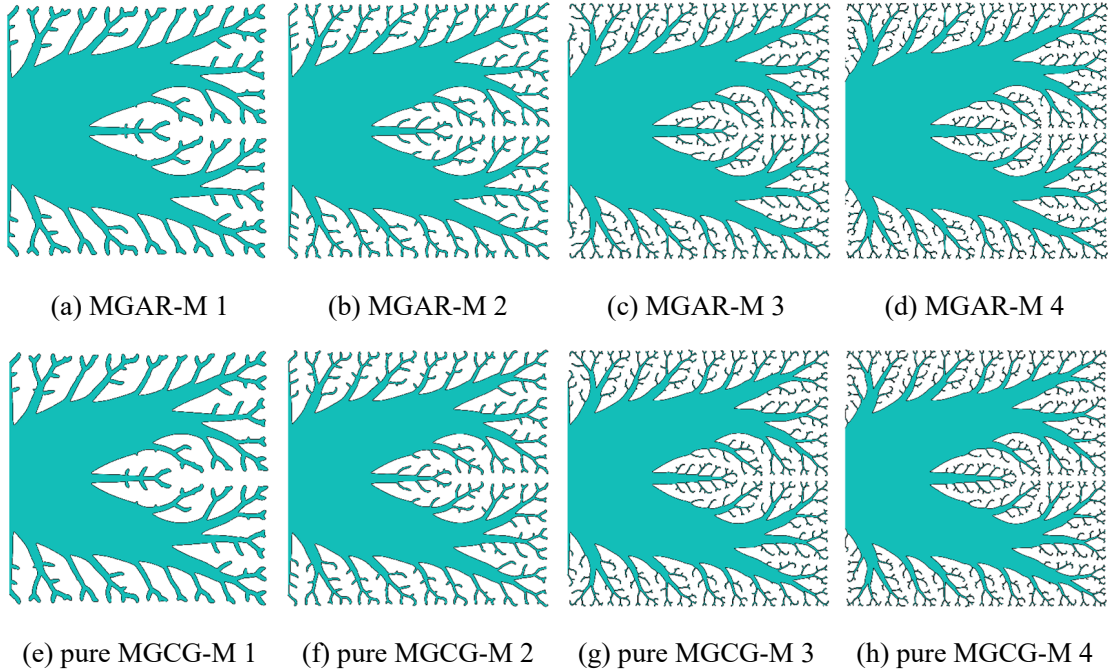

(a) MGAR-M 1　　　(b) MGAR-M 2　　　(c) MGAR-M 3　　　(d) MGAR-M 4

(e) pure MGCG-M 1　(f) pure MGCG-M 2　(g) pure MGCG-M 3　(h) pure MGCG-M 4

Figure 12 Structural topology of optimization result with different FE mesh scale.

It is well known that for topology optimization, close optimization objective functions can still be achieved even if there are significant differences in structural topology. Therefore, all optimized results are given in Figure 12 for comparison, the mesh scale of each structure in the Figure 12 increases from left to right (For simplicity, the "Model" is abbreviated as "M"). The results obtained by MGAR are almost indistinguishable from those obtained by pure MGCG, even though the details become more complex when dealing with large-scale problems. The "trunk" form of the structural topology does not change as the FE mesh scale increases, but simply grows more details to satisfy the structural thermal requirements. It should be noted that the concept of detail is only relative to the overall design, and as can be seen in Figure 13, even the thinnest part still guarantees more than 5 times the element length. Therefore, the post-processing strategy proposed in this work is reliable and achieve the purpose of improving the manufacturability of optimization results.

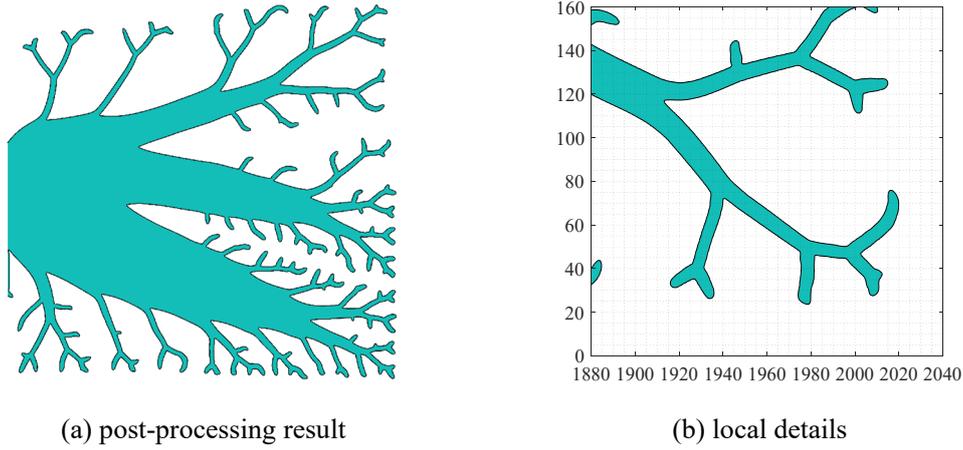

(a) post-processing result          (b) local details

Figure 13 Structural topology of optimization result with details for Model 4.

## 4.2 2D non-uniform heat source model

The 2D non-uniform heat source example is discussed in this section. The whole design domain is divided into four sub-domains with the same shape but different internal heat sources per unit volume, however, the internal heat sources inside each subdomain are uniform. As shown in Figure 14, $\Omega_1$ = 5e-5, $\Omega_2$ = 1e-4, $\Omega_3$ = 1.5e-4, $\Omega_4$ = 3e-4. The design domain is discretized with a 480 × 480 mesh and a 960 × 960 mesh respectively for the FEA of temperature field. The desired volume fraction was considered for both 0.4 and 0.6. The other boundary conditions and control parameters are consistent with the example in Section 4.1.

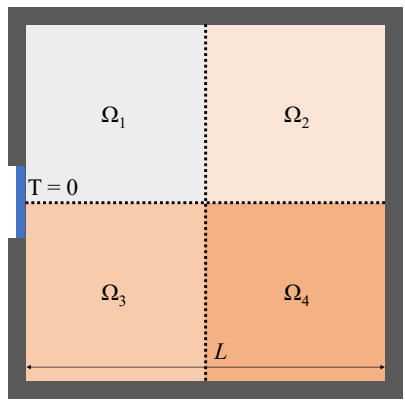

Figure 14 Design domain and boundary conditions for 2D non-uniform heat source model.

The average computational cost of MGAR with a volume fraction of 0.4 (performed on a 480 × 480 finite element mesh) is used as a unit cost measure, and the cumulative computational cost curves are given in Figure 15. For the non-uniform heat source problem, MGAR still maintains the advantage in terms of computational cost savings, and the computational efficiency is also at a close level compared to the uniform heat source problem. In addition, the increasing trend of the cumulative cost curve is also consistent with that in the uniform heat source problem, which implies that MGAR might be insensitive to the uniformity of the heat source distribution. It is a useful characteristic for the practical application of MGAR in engineering because non-uniform heat source scenarios are more common in nature.

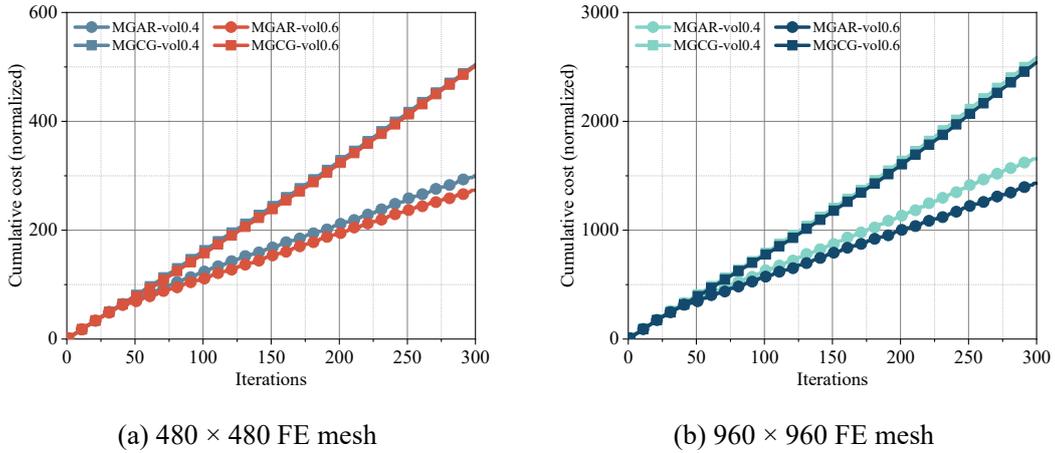

(a) 480 × 480 FE mesh  (b) 960 × 960 FE mesh

Figure 15 Normalized cost for 2D non-uniform heat source problems.

Table 3 Comparison of computational results of 2D non-uniform heat source model.

| scale | volume fraction | average CG iterations | MGCG evaluations | efficiency improvement | difference |
|---|---|---|---|---|---|
| 480 × 480 | 0.4 | 7.71 (7.38) | 112 | 40.50% | 0.13% |
| MGAR | 0.6 | 7.69 (7.32) | 102 | 45.46% | 0.0050% |
| 960 × 960 | 0.4 | 7.81 (7.51) | 124 | 35.93% | 0.028% |
| MGAR | 0.6 | 7.47 (7.24) | 109 | 43.68% | 0.0015% |

The influence in cost cannot be ignored when changing the volume fraction, and this is mainly evident for MGAR. However, changing the volume fraction has actually recast a new problem and it is reasonable to have a difference in computational cost, not to mention that increasing the volume fraction is beneficial to save computational cost. This phenomenon is even more remarkable when solving large-scale problems. The statistics in Table 3 show that the efficiency improvement with volume fraction of 0.6 up to 45.46%, which is a 4.96% increase relative to that with volume fraction of 0.4 on the 480 × 480 FE mesh. The same increase indicator reaches 7.75% on the 960 × 960 FE mesh. In other words, the influence of problem scale on efficiency improvement is diminished in the context of large volume fractions.

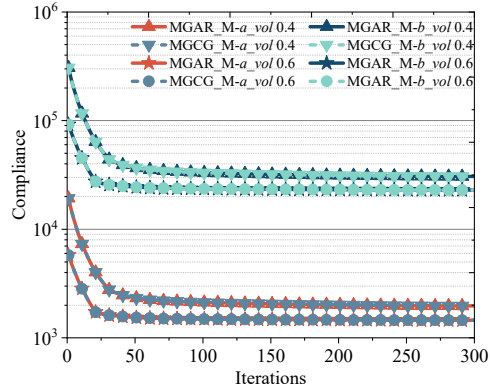

Figure 16 Compliance curves for different models, the M-*a* represent the 480 × 480 FE mesh and the M-*b* the 960 × 960 FE mesh.

The average CG iterations (the data in parentheses is associated with pure MGCG) and cumulative MGCG evaluations were counted in Table 3. First, the influence of volume fraction on the average CG iterations is very weak for both MGAR and pure MGCG, and the average CG iterations are close for the same volume fraction settings at different problem scales. Therefore, the average CG iterations is not the main reason for the difference in efficiency improvement caused by the volume fraction. Second, the cumulative MGCG evaluations called by MGAR is influenced by the volume fraction. Regardless of the problem scale, the MGCG evaluations decreased at large volume fraction settings, with only differences in the magnitude of the decrease. The gap of MGCG evaluations becomes larger when solving large-scale problems, which explains the difference in efficiency improvement caused by the volume fractions setting. In other words, a large volume fraction facilitates the reduction of the sensitivity of MGCG evaluations to problem scale. Final, the sensitivity of the MGCG evaluations is reflected in the efficiency improvement, which are less attenuated for large volume fraction settings when the problem scale increases. Moreover, it should be noted that the average CG iterations of MGCG in MGAR is larger than that of pure MGCG, so the efficiency improvement is attributed to MGAR. In summary, the volume fraction influences the computational efficiency of MGAR by influencing the MGCG evaluations, while it has almost no influence on pure MGCG, and a large volume fraction is more conducive to improve efficiency.

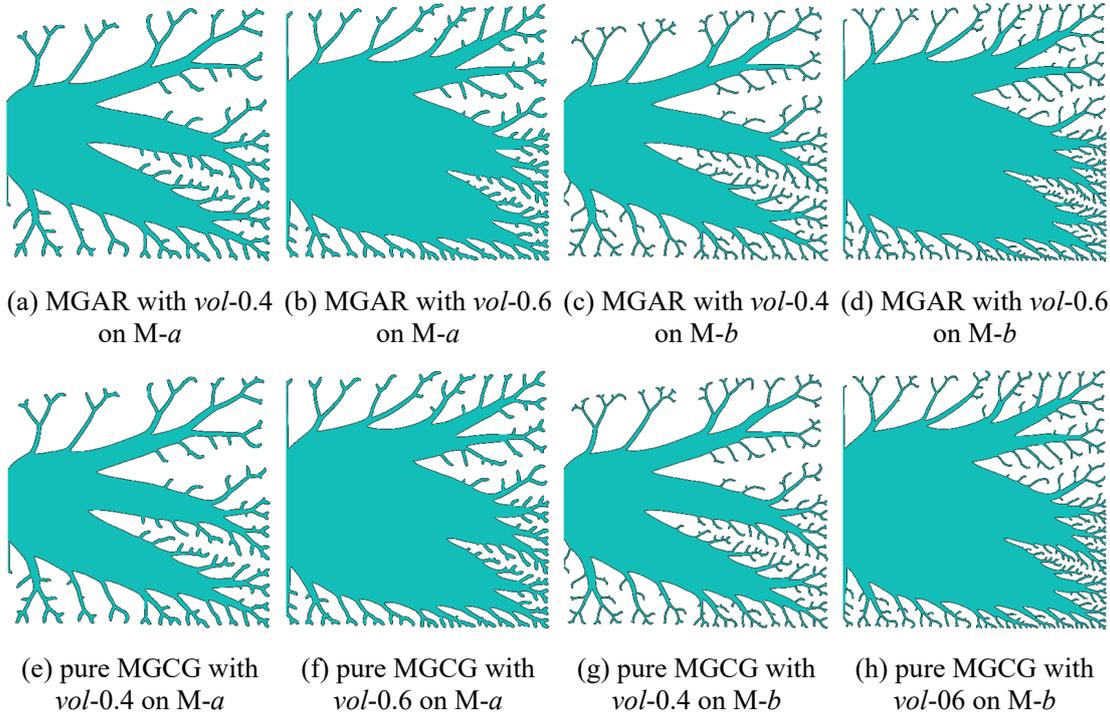

(a) MGAR with *vol*-0.4 on M-*a* (b) MGAR with *vol*-0.6 on M-*a* (c) MGAR with *vol*-0.4 on M-*b* (d) MGAR with *vol*-0.6 on M-*b*

(e) pure MGCG with *vol*-0.4 on M-*a* (f) pure MGCG with *vol*-0.6 on M-*a* (g) pure MGCG with *vol*-0.4 on M-*b* (h) pure MGCG with *vol*-06 on M-*b*

Figure 17 Structural topology of optimization results for 2D non-uniform heat source model.

A positive finding is that the influence of heat source distribution on the efficiency improvement and the consistency of optimization results so weak as to be negligible. As shown in Figure 16, for the whole optimization process with non-uniform heat source, the optimized objective curves of MGAR and pure MGCG with the same settings almost exactly coincident, and the maximum difference in the final optimization objective of MGAR relative to pure MGCG is also only 0.13%. The structural topology of the optimized results is given in Figure 17, it is clearly to see that it is difficult to distinguish the difference in details between the results obtained by the two methods, both for different problem scales and volume fraction setting scenarios.

### 4.3 3D uniform heat source models

To demonstrate the competency of MGAR, this section extends MGAR to 3D heat transfer applications. The whole design domain is a cube with side length $L$ as shown in Figure 18, the internal heat source per unit volume is set to 1e-4, and material properties of the structural model are set the same as in Section 3.2. The shaded area on the back denotes the Dirichlet boundary condition with T = 0, and its shape is a square with side length $l$ (one-eighth of $L$), and the remaining outer surfaces are adiabatic boundaries. The design domain is discretized with a 224 × 224× 224 FE mesh, however, only 1/4 model is taken for analysis due to the symmetry of design domain. The CARM consists of two basis vectors, the activation parameter $N_{on}$ is 2/15 of the maximum iteration, the volume fraction is 0.3, the minimum filter radius $r_{min}$ is 3, and the maximum number of CG iterations $cg_{max}$ is reset to 50.

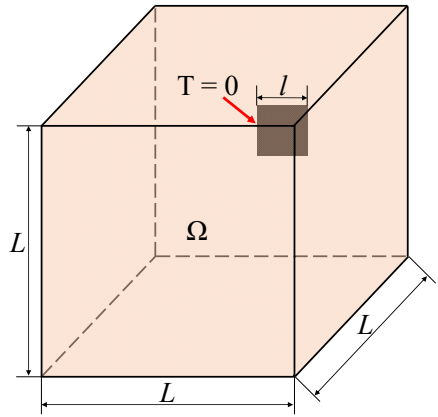

Figure 18 Design domain and boundary conditions for 3D uniform heat source model.

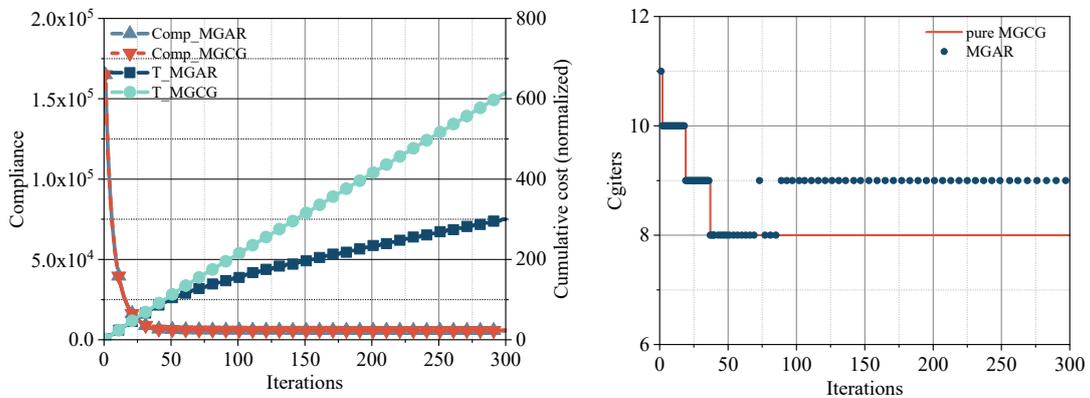

Figure 19 Compliance and cumulative cost curves for 3D uniform heat source model.

Figure 20 CG iterations for 3D uniform heat source model.

The MGAR is applied to 3D problem (DOFs of 2,873,025) to show even better performance, the efficiency improvement of MGAR over pure MGCG reaches 51.25%, much better than that of 2D uniform heat source problems with the approximate scale (efficiency improvement of the 2D model with DOFs of 2,436,721 is 35.18%). The MGAR has the potential to perform even better in terms of efficiency improvement in 3D scenarios. Figure 20 illustrates that the CG iterations of MGAR are all equal to or higher than the pure MGCG, and therefore the source of efficiency improvement is excluded from calling MGCG in MGAR. In addition, the average CG iteration of MGAR in the 3D scenario is close to the level in the 2D scenario, so the computational burden caused by the CG iterations is stable.

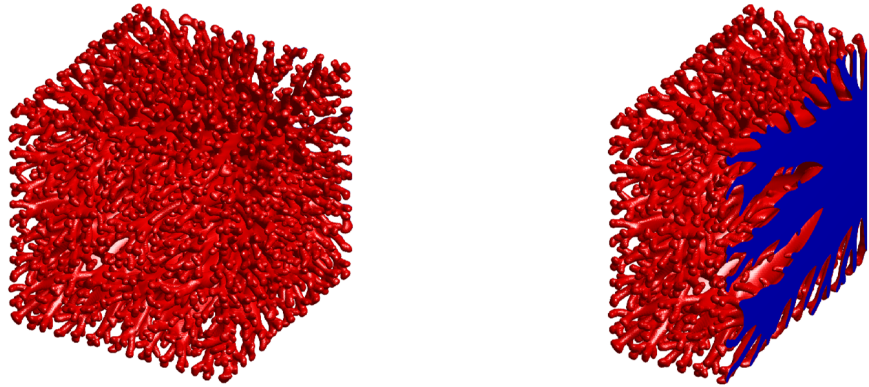

(a) MGAR (*obj* 5792.03)

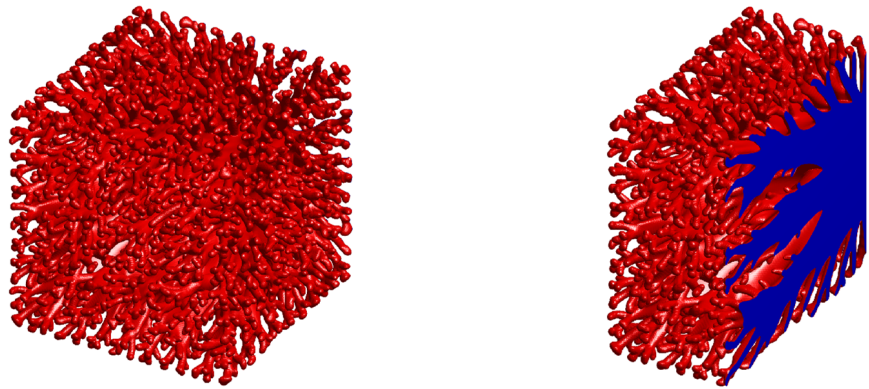

(b) MGCG (*obj* 5794.66)

Figure 21 Structural topology of the optimization results for 3D uniform heat source model.

The computational cost that are counted in Figure 19 is normalized, the actual computational cost of solving a 3D problem is much more expensive than that of a 2D problem with the same scale of DOFs. Therefore, the computational advantages of MGAR in 3D scenarios are excellent, both in terms of efficiency improvement and total computational cost savings. Furthermore, the optimization accuracy of MGAR in the 3D scenario is also guaranteed, the optimization objective curves almost completely coincident throughout the optimization process, and the relative difference of the final optimization objective is only 0.045%. The topology of the optimized structure is given in Figure 21, where the branches are coral-like and distributed throughout the design domain as much as possible. There is almost no detail difference between the optimized results of MGAR and pure MGCG, and the test results in this section show that MGAR is equally applicable and even performs better in 3D scenario.

## 5. Conclusion

This study proposes an efficient topology optimization method for heat transfer problems, named Multigrid Assisted Reanalysis (MGAR), which can relieve the burden of hardware to solve large-scale problems with significant computational cost savings. Moreover, a post-processing strategy is also proposed to improve the manufacturability of the results, and integrated with a continuous density filtering strategy to successfully obtain smooth boundary expressions.

Test results show that the proposed method is more efficient than MGCG. The efficiency improvement is weakened with the increase of scale for 2D problems, but the computational cost savings are increasing, the weakened efficiency improvement gradually saturate and might be stabilize at an acceptable value. The maximum efficiency improvement reaches 41.65%, and the efficiency improvement still reaches 34.10% on approximately 4.2 million DOFs. Moreover, MGAR is insensitive to the uniformity of heat source distribution and shows more excellent performance in 3D scenario. The efficiency improvement of MGAR over pure MGCG much better than that of 2D uniform heat source problems with the approximate scale. In addition to the excellent computational efficiency, the accuracy of optimization is also guaranteed, and the differences between optimization objective functions and structural topology are controlled at an almost negligible level.

The proposed post-processing strategy yields smooth boundary while ensuring consistent volume and material distribution before and after processing. Although the optimization objective becomes slightly worse after post-processing, this is acceptable considering the improved manufacturability of the optimization results. The application of continuous density filtering strategy in the post-processing strategy facilitates the elimination of some small-sized features and avoids discontinuous variations between topology optimization iterations.

# Ethics declarations

## Conflict of interest

The authors declared that they have no conflicts of interest with this work. We declare that we do not have any commercial or associative interest that represents a conflict of interest in connection with this work.

## Replication of results

An efficient heat transfer topology optimization, and post-processing strategy are described in detail in the paper. The corresponding algorithm is implemented in Matlab language. The authors are confident that the overall methodology can be reproduced. Due to the particularities of the project, the source codes and data cannot be shared. Readers are encouraged to contact the corresponding author.

# Acknowledgments

This work has been supported by the Program of National Natural Science Foundation of China under the Grant Numbers 11572120 and 51621004.